\documentclass{mem}
\usepackage{natbib}\usepackage{txfonts}\usepackage{balance}
\usepackage{graphicx}
\usepackage[a4paper]{hyperref}
\begin{document}

\title{AGN and Galaxy evolution from Deep X-ray surveys}

\subtitle{}

\author{
P. \,Tozzi\inst{1} \and R. Gilli\inst{2}
\and the CDFS Team}

\offprints{P. Tozzi}
 
\institute{INAF -- Osservatorio Astronomico di Trieste, via
G.B. Tiepolo 11, I--34131 Trieste, Italy \email{tozzi@ts.astro.it} \and
INAF -- Osservatorio Astronomico di Bologna, via Ranzani 1, I--40127,
Bologna, Italy \email{roberto.gilli@bo.astro.it}}

\authorrunning{P. Tozzi \& R. Gilli}

\titlerunning{Latest results from the CDFS}

\abstract{Deep X--ray surveys are providing crucial information on the
evolution of AGN and galaxies.  We review some of the latest results
based on the X--ray spectral analysis of the sources detected in the
Chandra Deep Field South, namely: i) constraints on obscured
accretion; ii) constraints on the missing fraction of the X--ray
background; iii) the redshift distribution of Compton--thick sources
and TypeII QSO; iv) the detection of star formation activity in
high--z galaxies through stacking techniques; v) the detection of
large scale structure in the AGN distribution and its effect on
nuclear activity.  Such observational findings are consistent with a
scenario where nuclear activity and star formation processes develop
together in an anti--hierarchical fashion.   } \maketitle{}

\section{Introduction}

In the last years deep X--ray surveys with the {\sl Chandra} and {\sl
XMM--Newton} satellites (Brandt et al. 2001; Rosati et al. 2002;
Alexander et al. 2003; Hasinger et al. 2001), paralleled by
multiwavelength campaigns (see, e.g., GOODS, Giavalisco et al. 2004),
provided several crucial information on the evolution of the AGN and
galaxy populations.  The bold result from the two deepest X--ray
fields, the Chandra Deep Field North (CDFN, observed for 2 Ms) and the
Chandra Deep Field South (CDFS, observed for about 1Ms) is constituted
by the resolution of the X--ray background (XRB) into single sources,
mostly AGN, at a level between 80\% and 90\% (see Bauer et al. 2004
and the recently revised estimate by Hickox \& Markevitch 2005),
providing an almost complete census of the accretion history of matter
onto supermassive black holes through the cosmic epochs.  However, the
most interesting outcomes go well beyond the demographic
characterization of the extragalactic X--ray sky.  Indeed, the
physical and evolutionary properties of the AGN population are now
revealing how they formed and how they are linked to their host
galaxies.  For the first time, the luminosity function of AGN has been
measured up to high redshift.  A striking feature is the {\sl
downsizing}, or {\sl anti--hierarchical} behaviour, of the nuclear
activity: the space density of the brightest Seyfert I and QSO is
peaking at $z\geq 2$, while the less luminous Seyfert II and I peak at
$z\leq 1$ (Ueda et al. 2003; Hasinger et al. 2005; La Franca et
al. 2005).  An analogous behaviour is presently observed in the cosmic
star formation history: at low redshift star formation is mostly
observed in small objects (see, e.g., Kauffmann et al. 2004), while at
redshift 2 or higher, star formation activity is observed also in
massive galaxies (with $M_* \sim 10^{11} M_\odot$, see, Daddi et
al. 2004a).

The global picture, as outlined by the present data, requires a tight
link between the formation of the massive spheroids and the central
black holes, as witnessed by the relation between black hole and
stellar masses or between black hole mass and the velocity dispersion
of the bulge (Kormendy \& Richstone 1995; Magorrian et al. 1998;
Ferrarese \& Merritt 2000).  The anti--hierarchical behaviour in both
star formation and AGN activity (which reflects in an
anti--hierarchical supermassive black holes growth, see Merloni 2004;
Marconi et al. 2004; for an alternative view see Hopkins et al. 2005),
is envisaged by theoretical models where energy feedback is invoked to
self--regulate both processes (see Fabian 1999; Granato et al. 2004).

In these Proceedings, we will describe a few observational results
obtained from the latest analysis of the X--ray and optical data in
the Chandra Deep Field South and North and which, in our view, are
consistent with this picture.  In detail, these results concern the
following issues:

\begin{itemize}
\item the physical properties of AGN from the X--ray spectral analysis
of faint X--ray sources;

\item the missing fraction of the XRB;

\item the distribution of obscured QSO and Compton--thick sources and
their relation with the cosmic mass accretion history;

\item star formation in high--z galaxies measured in the X--ray band
thanks to stacking techniques;

\item the effects of large scale structure onto nuclear activity.

\end{itemize}

\section{Properties of faint AGN from X--ray spectral analysis}

\begin{figure}[]
\resizebox{\hsize}{!}{\includegraphics{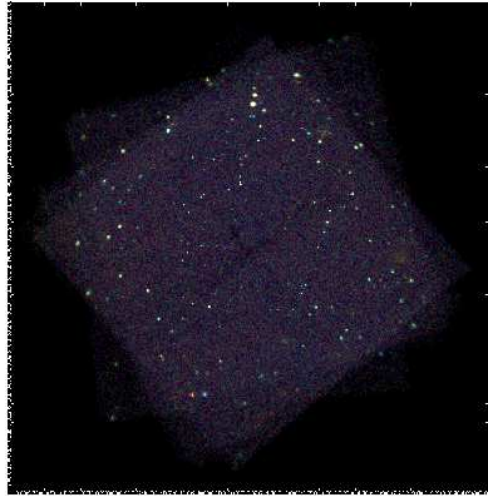}}
\caption{\footnotesize The color image of the CDFS (1 Ms exposure,
Rosati et al. 2002).  The X--ray color code is: red: 0.3--1 keV;
green: 1--2 keV; blue: 2--7 keV.  Note that fainter sources appear
more absorbed (bluer) than brighter ones. }
\label{fig1}
\end{figure}

\begin{figure}[]
\resizebox{\hsize}{!}{\includegraphics{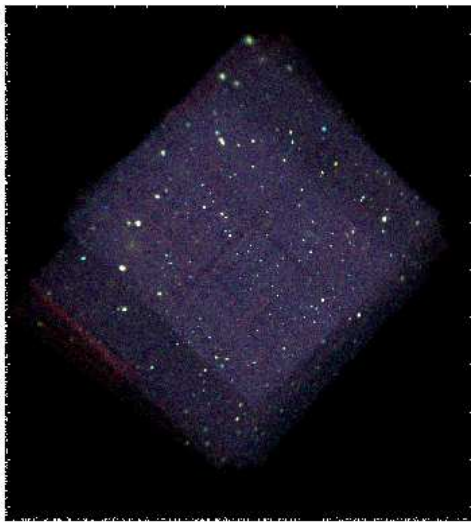}}
\caption{\footnotesize The color image of the CDFN (2 Ms
exposure). Same color code as of Figure \ref{fig1}.  }
\label{fig2}
\end{figure}

The resolution of the XRB, the long--awaited result since its
discovery in 1962, has been obtained simply by counting the point
sources found in the deep X--ray images taken with the {\sl Chandra}
and {\sl XMM--Newton} satellites.  This result is clearly shown by the
sharp images of the CDFS and the CDFN in Figure \ref{fig1} and
\ref{fig2}.  These images also show visually the solution of the
so--called {\sl spectral paradox}: fainter sources are more absorbed
(and appear bluer in X--ray colors) than brighter ones, so that the
total spectrum of the XRB, resulting from the summed contribution of
the whole AGN population, has a slope $\Gamma \simeq 1.4$, flatter
than that typical of the intrinsic nuclear emission $\Gamma = 1.8$ (as
observed in unabsorbed AGN).  The resolved fraction of the XRB has
been recently revised slightly downwards to be about 80\% in both
bands (Hickox \& Markevitch 2005).  AGN makes the 83\% and the 95\% of
the resolved fractions in the soft and in the hard band respectively.
On the other hand, star forming galaxies contributes only 3\% and 2\%
(Bauer et al. 2004).

However, quoting the resolved fraction of the XRB in the 2--8 keV band
is somewhat misleading.  Indeed, it has been pointed out that the
resolved fraction is significantly decreasing with increasing energy
(Worsley et al. 2005).  In particular, above 5 keV, the resolved
fraction can be as low as 50\%.  This finding opens again the issue of
the resolution of the XRB, requiring the presence of a still
undetected population of strongly absorbed AGN at moderate redshift,
as can be inferred from the spectral shape of the missing XRB (see
Worsley et al. 2005).

The issue of the missing XRB opens several questions on the physical
properties of the X--ray sources and calls for a detailed X--ray
spectral analysis of the faint AGN population.  In a recent Paper
(Tozzi et al. 2006) we went through a detailed X--ray spectral
analysis of the large majority of the X--ray sources found in the CDFS
(321 in the 1Ms catalog after excluding stars and low luminosity
sources with $L_X < 10^{41}$ erg s$^{-1}$, see Giacconi et al. 2002).
In particular, the knowledge of the spectroscopic or photometric
redshift for almost all the X--ray sources (Szokoly et al. 2004; Zheng
et al. 2004), allowed us to measure the value of the intrinsic
absorption in terms of equivalent hydrogen column density $N_H$.

\begin{figure}[]
\resizebox{\hsize}{!}{\includegraphics{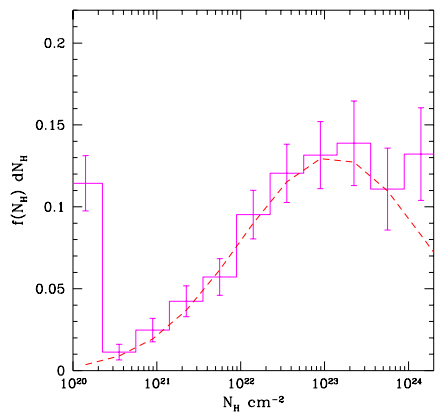}}
\caption{\footnotesize Intrinsic $N_H$ distribution representative of
the whole AGN population in the CDFS.  Errors are obtained from the
poissonian uncertainties on the number of detected sources in each
bin.  The dashed curve is a lognormal distribution with $\langle
log(N_H) \rangle = 23.1$ and $\sigma = 1.1$.  All Compton--thick
candidates have been placed in the bin at $N_H = 10^{24}$ cm$^{-2}$
(Tozzi et al. 2006). }
\label{nh}
\end{figure}

Summarizing the main results of our analysis, we found that the
intrinsic spectral slope $\Gamma$ is always close to the average value
$\Gamma=1.8$, without showing any significant dependence on redshift,
intrisic luminosity, or intrinsic absorption.  We find significant
evidence of the 6.4 keV Fe line in 12\% of the sources with
spectroscopic redshift.  We detect the presence of a soft component
(possibly due to partial covering or to scattered emission) in only 8
sources.  We measured the intrinsic column density $N_H$ for each
source, after freezing the spectral slope $\Gamma=1.8$ for the
faintest ones.  The intrinsic $N_H$ distribution has been obtained
after correcting for the detection probability as a function of the
flux (the sky--coverage) and for sources below the limiting flux of
the survey.  We find that most of the AGN have high intrinsic
absorption with $N_H > 10^{22}$ cm$^{-2}$ (see Figure \ref{nh}).  A
fraction of the AGN (more than 10\%) are Compton--thick sources,
defined as sources with $N_H \geq 1.5 \times 10^{24}$ cm$^{-2}$.  Only
few of these elusive sources are actually detected (we identify only
14 Compton--thick candidates in the CDFS), buth their actual number
density is expected to be high.  Indeed, due to their spectral shape,
well represented by a very hard reflection spectrum, only a small part
of their population can be detected by {\sl Chandra}, which is mostly
sensitive in the soft band.  We also find that about 80\% of the
sources (among the 139 AGN for which good optical spectra are
available), follow a one--to--one correspondence between optical Type
I (Type II) and X--ray unabsorbed (X--ray absorbed, defined as sources
with $N_H \geq 10^{22}$ cm$^{-2}$) sources, as predicted by the
original version of the unification model for AGN (Antonucci 1993).

\section{The missing fraction of the XRB}

The detailed knowledge of the spectral shape allows us to better
evaluate the detection probability of each source class.  This is an
important aspect, since the probability of being included in a survey
depends significantly on the spectral shape.  Therefore, by a careful
spectral analysis, we are able to estimate with high accuracy the
contribution of the most elusive absorbed sources to the XRB.  Indeed,
the detection probability is lower for more absorbed sources, since
the maximum detectability is achieved in the soft band; therefore, the
contribution to the XRB of strongly absorbed sources found in the CDFS
is higher than that of sources with the same energy flux but with no
or little absorption.  This aspect was not fully appreciated in
previous works, where the detection probability was only a function of
the flux and not of the spectral shape, with a consequent
underestimate of the actual number density of the most absorbed
sources.

\begin{figure}[]
\resizebox{\hsize}{!}{\includegraphics{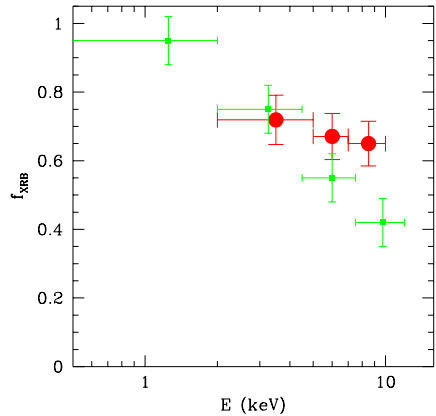}}
\caption{\footnotesize Red circles: contribution to the resolved XRB
in the CDFS in different energy bands.  Small green squares are from
Worsley et al. (2005). }
\label{xrb}
\end{figure}

We recompute the resolved XRB in the CDFS, and compare it to the total
extragalactic XRB spectrum modeled as a power law with $\Gamma = 1.41$
and 1 keV normalization of 11.6 keV cm$^{-2}$ s$^{-1}$ sr$^{-1}$
keV$^{-1}$ (De Luca \& Molendi 2004).  The contribution to the XRB
from CDFS sources is obtained directly by summing the contributed flux
from each source weighted by the inverse of the detection probability.
The contributed flux in a given energy band is obtained by fitting the
X--ray spectrum of each source with a power law plus an intrinsic
absorption.  We find that the decrease of the resolved fraction as a
function of the energy range is not as pronounced as in Worsley et
al. (2005), as shown in Figure \ref{xrb} (Tozzi et al., in
preparation).  This implies that we are actually seeing a fraction of
the population responsible for the missing XRB.  Due to the strong
absorption, most of these sources can not be detected even in the
deepest {\sl Chandra} or {\sl XMM--Newton} surveys, and their
discovery must wait for an higher energy, high--sensitivity X--ray
mission, or make use of radio or submillimetric data (see
Mart\`inez--Sansigre et al. 2005).

\section{AGN and Galaxy formation: TypeII QSO and Compton--thick 
sources}

\begin{figure}[]
\resizebox{\hsize}{!}{\includegraphics{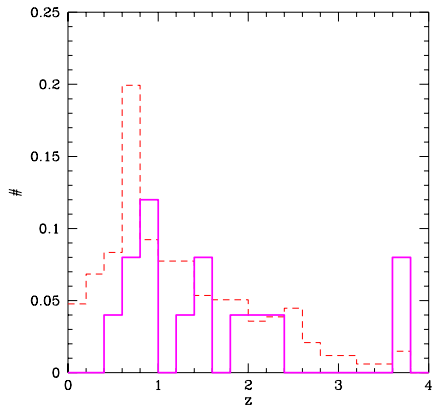}}
\caption{\footnotesize Normalized redshift distribution of the
Compton--thick candidates in the CDFS (14 sources, solid histogram)
compared with the normalized distribution of the whole sample (321
sources, dashed histogram).}
\label{cthick}
\end{figure}

Another interesting piece of information comes from the redshift
distribution of the Compton--thick candidates in the CDFS.  As shown in
Figure \ref{cthick}, these sources are distributed in a wide range of
redshift, a significant part of them around $z\sim 1$.  Their redshift
and their level of absorption match well with the values expected for
the sources responsible of the missing XRB (see Worsley et al. 2005),
reinforcing the reliability of our candidate sources.

\begin{figure}[]
\resizebox{\hsize}{!}{\includegraphics{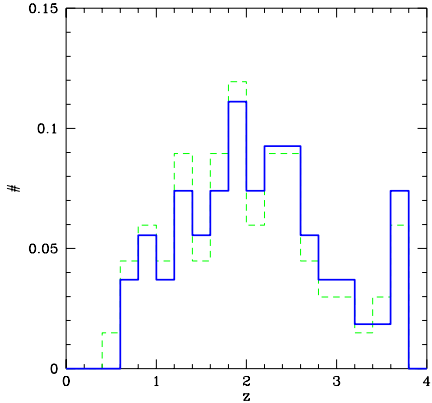}}
\caption{\footnotesize Normalized redshift distribution of the
absorbed QSO in the CDFS (54 sources, solid histogram) compared with
the normalized distribution of all the sources with intrinsic X--ray
luminosity $L > 10^{44}$ erg s$^{-1}$ (67 sources, dashed histogram)}
\label{qso2}
\end{figure}

Another interesting class of sources are the so called Type II QSO.
Since we do not have an optical spectral classification for all the
sources, we consider here absorbed QSO, simply defined on the basis of
the X--ray properties as bright sources ($L>10^{44}$ erg s$^{-1}$)
with $N_H>10^{22}$ cm$^{-2}$.  Some of them have been shown to
correspond to Type II QSO on the basis of optical and submm data
(Norman et al. 2002, Mainieri et al. 2005a).  Most of them are among
the optically faint sources of the sample (Mainieri et al. 2005b).  We
select 54 sources with these properties distributed on a wide range of
redshifts (see Figure \ref{qso2}), corresponding to 80\% of the
sources with $L>10^{44}$ erg s$^{-1}$ in our sample.  We remark that
we explore here a limited luminosity range ($L < 10^{45}$ erg
s$^{-1}$), given the small volume sampled.  This confirms anyway that
Type II QSO constitute a significant fraction of the AGN population.

These findings can be interpreted in the framework of the
anti--hierarchical scenario, as described in the Introduction.
Absorbed QSO may be sources associated to massive spheroids
experiencing at the same time rapid growth of the central black hole
and strong star formation activity.  In this case, the absorption is
not ascribed to circumnuclear matter, as in the simplest version of
the unification model, but to the gas distributed on a much wider
region strongly polluted by star formation processes.  Indeed, the
redshift distribution of the QSO population, and therefore of the
absorbed ones which represent a significant fraction of it, peaks at a
redshift $\geq 2$, an epoch when the number density of massive and
powerful starburst galaxies is expected to be interestingly high, as
found recently with near--IR observations (Daddi et al. 2004, see \S
5).

The strong energetic feedback from both nuclear activity and star
formation in such large objects, is expected to inhibit further
accretion and star formation events.  On the other hand, smaller
objects, where the feedback is less efficient, are able to retain gas
that can be accreted subsequently, allowing for episodes of obscured
accretion and star formation at lower redshifts. 

\section{Star formation seen in X-ray at high--z}

X--ray emission witnessing star formation events is due to X--ray
binaries and hot gas associated with superwinds and SNa remnants.  The
main advantage of X--ray studies of the cosmic star formation rate is
to avoid problems of obscuration, which severely affects optical
observations.  The price to pay is that, since star forming galaxies
have luminosities in the range $10^{39}$--$10^{42}$ erg s$^{-1}$, it
is difficult to detect them at high redshifts even in the deepest
X--ray surveys.  The first normal star--forming galaxy X--ray
luminosity function has been derived by Norman et al. (2004) in the
combined Chandra Deep Field North and South.  The results show an
increasing cosmic star formation rate proportional to $(1+z)^{2.7}$,
consistent with other star formation determination in different
wavebands.  However, the galaxy XLF is determined only at $z \leq 1$,
still below the expected maximum of the cosmic star formation history.

\begin{figure}[]
\resizebox{\hsize}{!}{\includegraphics{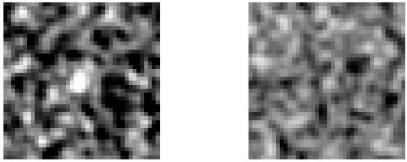}}
\caption{\footnotesize Stacked X--ray images in the soft (left) and
hard (right) band of 22 BzK selected galaxies in the CDFS field (see
Daddi et al. 2004b).}
\label{z2sb}
\end{figure}

On the other hand, other selection techniques, like the one using the
B--z vs z--K color diagram (see Daddi et al. 2004b), are able to
identify actively star forming galaxies at $z\geq 1.4$, with typical
stellar masses larger than $10^{11} M_\odot$ and average star
formation rate of $\simeq 200 M_\odot$ yr$^{-1}$.  The estimated
number density of these $z\sim 2$ star forming galaxies suggests that
we are peering into the formation epoch of massive early--type galaxies
(Daddi et al. 2004a).

The expected X--ray emission from each one of these high--z, starburst
galaxies is below the flux limits of the deepest X--ray surveys.
However, we can add together the images of the X--ray fields in the
positions of the galaxies, to obtain a stacked image of all the
optically selected star forming galaxies.  We created the stacked
images of 22 {\sl BzK} selected galaxies in the CDFS in the soft and
the hard bands.  This can be considered an image of about 20 effective
Ms of a typical $z\sim 2$ massive, star forming galaxy.  The images
are shown in Figure \ref{z2sb}.  In the soft band we detect a total
$96 \pm 23$ net counts, which, for an average spectral slope of
$\Gamma = 2.1$, corresponds to an average 2--10 keV rest frame
luminosity of $9 \times 10^{41}$ erg s$^{-1}$, impliying a star
formation rate of about $\sim 190 M_\odot$ yr$^{-1}$ (see Ranalli et
al. 2003), in good agreement with the estimate from the
reddening--corrected UV luminosities.  On the other hand, we have no
detection in the observed--frame hard band, confirming the steep
spectral slope (hardness ratio $HR < -0.5$ at the 2$\sigma$ level) and
therefore the non--AGN nature of these sources (Daddi et al. 2004b).

This findings confirm that stacking techniques on sources selected in
other wavebands, are extremely useful in exploring the level of X--ray
emission from star formation at high--z, an aspect which constitutes
one of the most compelling scientific cases for the next generation
X--ray facilities.

\section{Large Scale Structure in deep X--ray surveys}

\begin{figure}[]
\resizebox{\hsize}{!}{\includegraphics{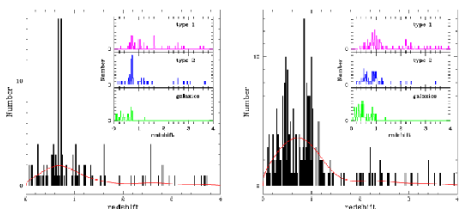}}
\caption{\footnotesize Redshift distribution of X--ray sources in CDFS
(left) and CDFN (right, Gilli et al. 2005).}
\label{gilli1}
\end{figure}

With the spectroscopic follow--up of X--ray sources detected in the
CDFS, significant large scale structure has been discovered both in
CDFS and CDFN, as shown in Figure \ref{gilli1}.  In particular, two
prominent spikes have been found in the CDFS at $z=0.67$ and $z=0.73$
with 19 sources each (Gilli et al. 2003), corresponding to spikes
found in the distribution of galaxies in the K20 survey (an ESO--VLT
optical and near--infrared survey down to $K \leq 20$ covering part of
the CDFS, see Cimatti et al. 2002).  Comparing the X--ray and optical
catalogs, we found that in the structure at $z=0.73$, the fraction of
active galaxies is the same as in the field, while in the one at
$z=0.67$ it is higher by a factor of 2.  We also note that the
structure at $z=0.73$ includes a cluster of galaxies, which may have
biased downwards the estimate of the AGN fraction.  This finding
constitutes one of the first tantalizing hints (significant only at
the 2$\sigma$ level) that large scale structure can trigger nuclear
activity.

\begin{figure}[]
\resizebox{\hsize}{!}{\includegraphics{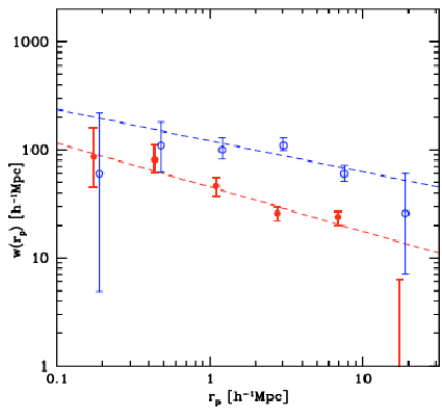}}
\caption{\footnotesize Projected correlation functions of X--ray sources
in the CDFS (open circles) and CDFN (filled circles, Gilli et
al. 2005).}
\label{gilli2}
\end{figure}

\begin{figure}[]
\resizebox{\hsize}{!}{\includegraphics{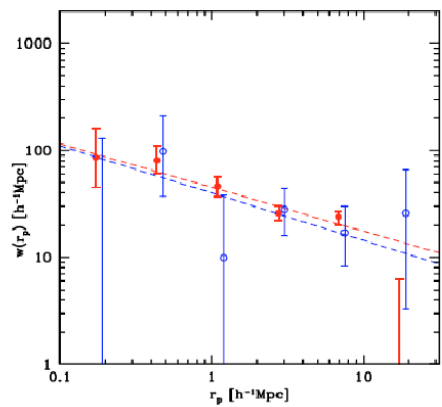}}
\caption{\footnotesize Spatial correlation functions in the CDFS
(open circles) after removal of the two major spikes compared with
that measured in the CDFN (filled circles, Gilli et al. 2005).}
\label{gilli3}
\end{figure}

Investigation of the spatial clustering of X--ray sources in both
fields (Gilli et al. 2005) points out a significant difference in the
correlation lengths.  We find $r_0 = 8.6 \pm 1.2 \, h^{-1}$ Mpc in the
CDFS, and $r_0 = 4.2 \pm 0.4 \, h^{-1}$ Mpc in the CDFN, with
similarly flat slope ($\gamma = 1.33 \pm 0.11$ and $1.42 \pm 0.07$
respectively), as shown in Figure \ref{gilli2}.  If we consider only
AGN, we obtain higher correlation lenghts, in the range $5-10 \,
h^{-1}$ Mpc.  Since at $z\sim 1$ late--type galaxies have a
correlation lenght of $\sim 3.2 \, h^{-1}$ Mpc while early--type
galaxies have $\sim ~6.6 \, h^{-1}$ Mpc (Coil et al. 2004), the high
correlation lengths measured for AGN in the CDFS are consistent with
the idea that at $z\sim 1$ AGN with Seyfert--like luminosities are
hosted by massive galaxies.  The difference in the correlation lengths
measured between the two fields disappears when the two most prominent
spikes in the CDFS are removed (see Figure \ref{gilli3}).  This shows
that larger fields of view are needed to kill the cosmic variance and
perform a proper investigation of the correlation properties of X--ray
detected AGN.  We mention two main projects, the COSMOS survey with
the {\sl XMM--Newton} satellite (Hasinger et al. in preparation), and
the Extended CDFS with {\sl Chandra}.  The Extended CDFS complements
the original 1Ms exposure with four {\sl Chandra} ACIS-I fields with
250 ks each, bringing the linear size of the field from the former 16
arcmin to 32 arcmin.  First results have been published by Lehmer et
al. (2005), while spectroscopic follow up is currently under way.  The
main goal is to better understand the effect of large scale structure
onto nuclear activity, and the evolution of the clustering of X--ray
sources.

\section{Conclusions}

We presented few selected topics which we find particularly relevant
among the latest results from deep X--ray surveys.  We can summarize
our conclusions as follows:

\begin{itemize}

\item a population of strongly absorbed, possibly Compton--thick AGN at
$z\sim 1$ is still missing to the census of the X--ray sky; the
detailed X--ray spectral analysis of faint sources shows that we are
detecting some of them, and help us in obtaining a complete
reconstruction of the cosmic accretion history onto supermassive black
holes;

\item we find several absorbed sources (the so--called TypeII QSO)
among the population of bright AGN, possibily witnessing the rapid
growth of the super massive black holes associated to strong star
formation events;

\item thanks to stacking techniques, we detected the X--ray emission
associated to massive star forming galaxies at redshift as high as
$z\sim 2$, therefore peering with X--rays in the epoch of massive
galaxy formation;

\item investigation of large scale structure in the X--ray detected
AGN distribution provides tantalizing hints of its effect on nuclear
activity.  Studies of spatial correlation of X--ray sources require
larger fields of view to kill the cosmic variance and therefore
evaluate properly the evolution of the AGN clustering properties.

\end{itemize}
 
Such observational findings are providing crucial information on the
evolution of AGN and galaxies.  Present--day data are consistent with
a scenario where nuclear activity and star formation processes
develop together in an anti--hierarchical fashion.

\begin{acknowledgements}

P.T. thanks the Organizers for providing a pleasant and stimulating
scientific environment during the workshop.

\end{acknowledgements}

\bibliographystyle{aa}

\end{document}